# Temporal changes of near-surface air temperature in Poland for 1781-2016 and in Tbilisi (Georgia) for 1881-2016

R. Modzelewska[1], M. V. Alania[1,2], N. I. Kapanadze[3], and E. I. Khelaia[2]

1. Institute of Mathematics and Physics, Siedlce University, Siedlce, Poland

2. M. Nodia Institute of Geophysics Ivane Javakhishvili Georgia State University, Tbilisi, Georgia

3. Institute of Hydrometeorology, Tbilisi, Georgia

Corresponding author: Renata Modzelewska   email: renatam@uph.edu.pl

Key Points:

- We analyze temporal changes of the near-surface air temperature in Poland and in Tbilisi (Georgia) in relation to solar activity.

- In global warming decisively contributes the period 1890-1960 when correlation between solar activity and the air temperature is very high.

- We identify disturbances in the near-surface air temperature data: ~20 ± 3 years and ~7-8 years.

- **In global warming decisively contributes the period 1890-1960 when correlation between solar activity and the air temperature is high.**

Abstract

Analyses of near-surface air temperature T in Poland for 1781-2016 and in Tbilisi (Georgia) for 1881-2016 have been carried out. We show that the centenary warming effect in Poland and in Tbilisi has almost the same peculiarities. An average centenary warming effect ΔT = (1.08 ± 0.29) C is observed in Poland and in Tbilisi for 1881-2016. A warming effect is larger in winter season (ΔT = ~1.15 C) than in other seasons (average warming effect for these seasons ΔT = ~0.95 C). We show that a centenary warming is mainly related to the change of solar activity (estimated by sunspot numbers (SSN) and total solar irradiance (TSI)); particularly, a time interval about ~70 years (1890-1960), when a correlation coefficients between 11 years smoothed SSN and T, and TSI and T are high, r = 0.66 ± 0.07 and r = 0.73 ± 0.07 for Poland and r = 0.82 ± 0.05 and r = 0.90 ± 0.05 for Tbilisi, respectively; in this period solar activity contributes decisively in the global warming. We show that a global warming effect equals zero based on the temperature T data in Poland for period 1781-1880, when human activities were relatively less than in 1881-2016. We recognize a few feeble ~ 20 ± 3 years disturbances in the temperature changes for period 1885-1980, most likely related with the fluctuations of solar magnetic cycles. We distinguish the fluctuations of ~7-8 years in Poland's T data, possibly connected with local effects of the North Atlantic Oscillation.

1. Introduction

A study of a global warming effect (centenary changes of air temperature T) is one of the greatest mankind's problem at present. In spite of much effort in many studies, unfortunately, there are not found any objective reasons of these changes; however in different publications one can find the discussion about the roles of the solar activities and/or human activities in these processes [e.g. *Solomon, et al.,* 2019; *Zherebtsov, et al.,* 2019; *Booth*, 2018; *Dorman*, 2016; 2004].
During the ~last half century, it is observed an increase in global temperature which causes the extensive discussion devoted to air temperature trends in the context of global warming [e.g. *Pittock*, 1979; *Grey et al.,* 2010; *Myhre et al.,* 2013]. Many of the papers suggest that the global warming in recent years is mostly caused by anthropogenic effects connected with increase of greenhouse gas concentrations in the air [e. g. *Barnett et al.,* 2001; *Myhre et al.,* 2013], others are against [e.g. *Easterbrook*, 2016]. However, the great majority of researchers suggest that solar activity plays a substantial role of global temperature variability [e.g. *Haigh*, 1996; 2007; *Svensmark,* 2007; *Gervais,* 2016; *Mendoza et al.,* 2019]. Currently, there are numerous arguments suggesting that the solar variability affects the global climate in different aspects and on different timescales [*Usoskin and Kovaltsov*, 2006]. There are two broad classes of mechanisms of forcing solar energy [*Grey et al.,* 2010]: via direct changes in total solar irradiance [for a review see, *Lockwood,* 2012] or by solar wind driven energetic particles affecting the cloud formation [for a review see, *Marsh and Svensmark,* 2003]. *Grey et al.* [2010] suggested that: "in both of these cases the forcing is likely to be very small. However, even a very weak forcing can cause a significant climate effect if it is present over a long time or if there are nonlinear responses giving amplifying feedbacks." Nevertheless, up to now the full physical mechanism for these aspects has been still the major challenge.
Thus, it could be underlined that any kind of study using experimental data not employed before for these purposes is definitely important; e.g from this point of view data of the near-surface air temperature in Tbilisi, provided by Institute of Hydrometeorology in Tbilisi (Georgia) are not published before at all.
Our aim in this paper is several fold : (1) to analyze long-period data of the near-surface air temperature (T) in Tbilisi (1881-2016) and in Poland (1781-2016) in relation with solar activity in order to establish, whether the changes of T shows a global warming effect, (2) to demonstrate that changes of T for this period can be divided in two sub time intervals- the first (1890-1960) with a strong positive correlation between solar activity data and T, and the second interval (1960-2016) with a relatively weaker negative correlation between them, (3) to show that the centenary change (global warming effect) generally falls to the winter season, and (4) to reveal some peculiarities, e.g. a tendency of quasi periodic 20 ± 3 years and 8 ± 2 years cycling of the T for the time interval presumably related with the solar activity and solar magnetic polarity.

2. Data analyzes

For analyzes we use annual data of the near-surface air temperature T in Poland for 1781-2016, in Georgia (observatory in Tbilisi) for 1881-2016 and global temperature T [*Hansen et al.*, 2010] for 1881-2016. The historical reconstructed near-surface air temperature T in Poland comes from different stations: since 1781 Warsaw [*Lorenc*, 2000], since 1791 Wroclaw [*Brys and Brys*, 2010], since 1825 Cracow [*Piotrowicz*, 2007], since 1837 Görlitz (Zgorzelec) (source Deutscher Wetterdienst), since 1848 Poznań [*Kolendowicz et al.*, 2019], since 1851 Gdansk and Hel [*Miętus*, 1998], since 1871 Toruń [*Pospieszyńska and Przybylak*, 2019], since 1881 Sniezka [*Głowicki*, 1998], since 1903 Lodz [*Wibig et al.*, 2004]. The actual near-surface air temperature T in Poland and the list of stations are provided by The Institute of Meteorology and Water Management - National Research Institute (IMGW-PIB) (https://dane.imgw.pl). In Poland the average daily air temperature was calculated as [*Urban*, 2010]: Tav=(T06 + T12 + 2 * T20)/4 until January 1, 1966 and from that date Tav=(T00 + T03 + T06 + ... + T21)/8. Data of temperatures in Tbilisi are recorded at the Hydrometeorological Observatory in Tbilisi (Georgia). The monitoring was conducted through the whole year (12 months), in 1881-1935 it was recorded three times a day at 7:00, 13:00 and 21:00, in 1936-1960 four times a day at 1:00, 7:00, 13:00, 21:00. Since 1966 till today monitoring is executed 8 times every day. The hours are given in UTC (GMT). According to these information, the monthly and annual average temperatures are calculated according to World Meteorological Organization (WMO) standards. As an indicator of solar activity we use the data of sunspot number (SSN) [*Balogh et al.*, 2014] and total solar irradiance (TSI) [*Coddington et al.*, 2016] for this period.

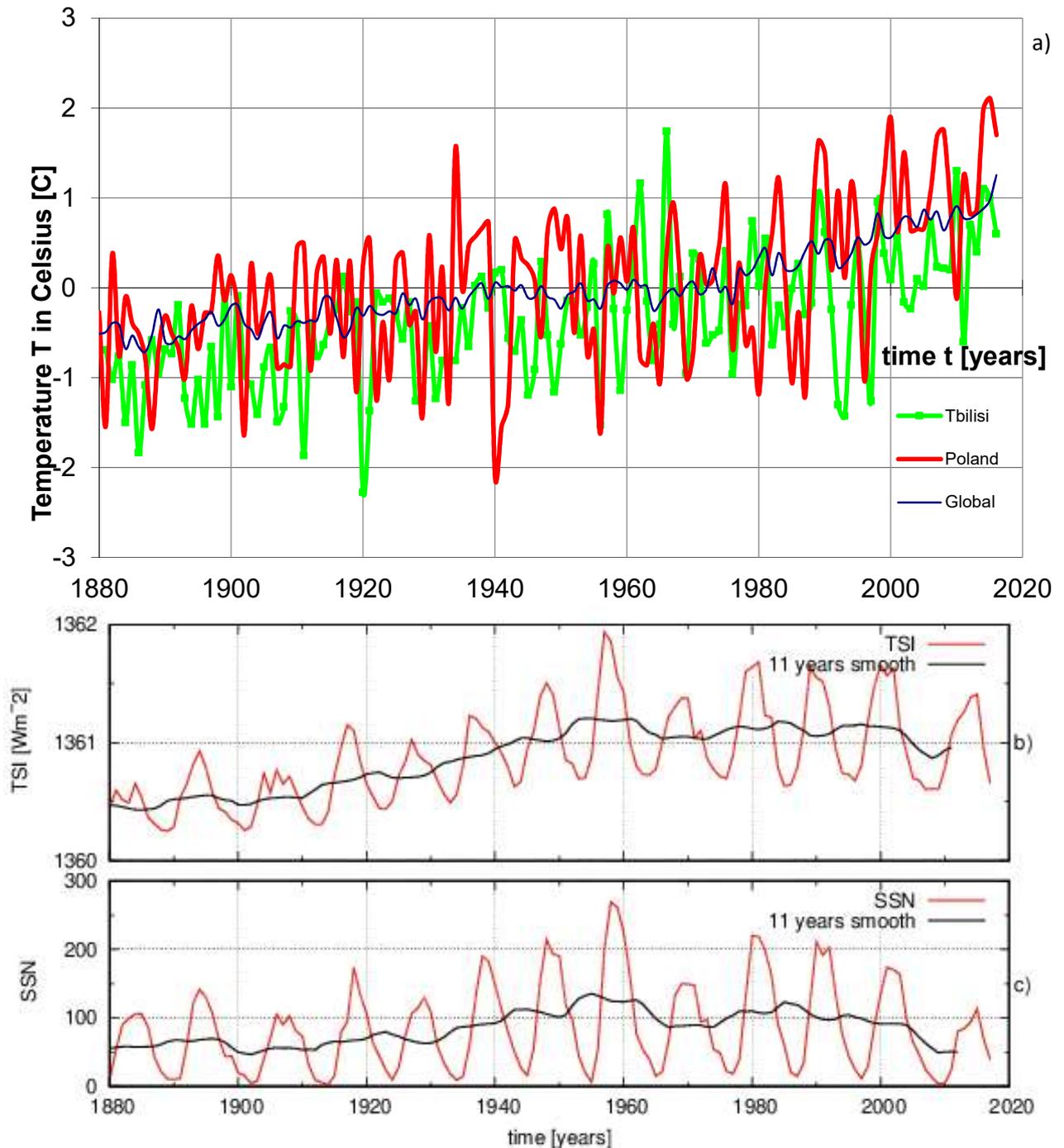

Figure 1abc. Temporal changes of the (a) annual near-surface air temperature T in Poland (red), Tbilisi (green) and global air temperature (black); (b) annual TSI and (c) SSN in 1881-2016.

In Figure 1a are presented annual changes of the near-surface air temperature T in Poland, Tbilisi and global temperature for 1881-2016. Figure 1bc show the annual changes of SSN and TSI for 1881-2016, indicating that ~ 90% of total energy of changes of SSN and TSI falls to the 11-year periodicities connected with solar activity cycle. Figure 1a displays a little non significant gradually increase (positive drift) with some fluctuations versus time t. The global T illustrates the average change of the near-surface air temperature around the world without short period oscillations, whereas the local measurements of temperature (e.g. in Poland and in Tbilisi) show much more variable structure with peculiar character on the shorter time scales. In this regard from one side it is essential to compare the local measurements of T with global temperature in centenary scales, confirming the relevance of local measurements; on the other hand the local near-surface air temperature data allows to analyze the high frequency oscillations. Thus, this approach gives the unique possibility to look for new features of the near-surface air temperature T in both, long and short period time scales. We assume that, a study of the features of the temporal changes of the local T measurements can be considered as an important aim of this paper, as well.

To reveal centenary changes of time series in T data, we use the least squares method of linear regression, T=at+b. The regression lines are drawn in Figures 2abc for Poland (a), Tbilisi (b) and global temperature (c) for 1881-2016. In horizontal (X) axis are presented years; in vertical axis (Y) is presented T in Celsius.

The regression lines show that a warming effect (centenary change) of T is $\Delta T=(1.11 \pm 0.33)^0C$ / per 100 years for Poland, $\Delta T=(1.04 \pm 0.25)^0C$ / per 100 years for Tbilisi and $\Delta T=(0.99 \pm 0.07)^0C$ / per 100 years for global T measurements for 1881-2016. So, the average warming effect for Tbilisi and Poland is $\Delta T=(1.08 \pm 0.29)^0C$ / per 100 years, this is in accord with previous study based on global temperature T data [ e.g. *Balling et al*. 1998; *Osborn et al.*, 2017].

It is well known that the greatest warming occurs in winter seasons, and the least warming occurs in summer seasons [e.g., *Jones et al.,* 1997; *Balling et al*., 1998]. It is of interest whether the centenary changes of T is identically evident in different seasons of year for Poland and Tbilisi T data. For this purpose we analyze data of T for hot months (April-May-June-July-August-September) and cold (January-February-March-October-November-December), separately. In the Figures 2def are presented annual temporal changes of above mentioned averaged over six months data of T for Poland (Figure 2d), Tbilisi (Figure 2e) and for global temperature changes (Figure 2f) for the period of 1881-2016 with corresponding regression lines. Figures 2def show that the centenary changes of near-surface air temperature in cold season ($\Delta T = \sim 1.2^0C$ for Poland, $\Delta T = \sim 1.1^0C$ for Tbilisi and $\Delta T = \sim 1^0C$ for Global) are larger than in summer season ($\Delta T = \sim 1^0C$ years for Poland, $\Delta T = \sim 0.9^0C$ for Tbilisi and $\Delta T = \sim 0.95^0C$ for Global).

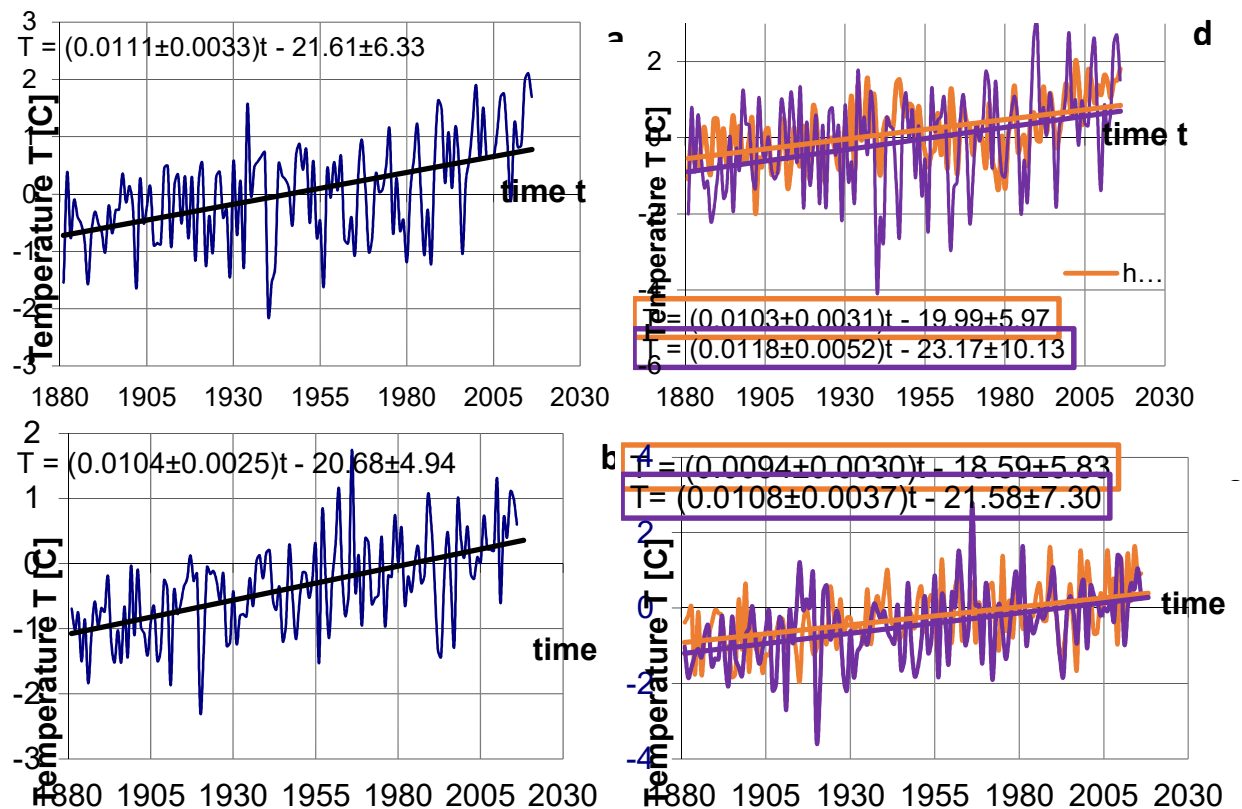

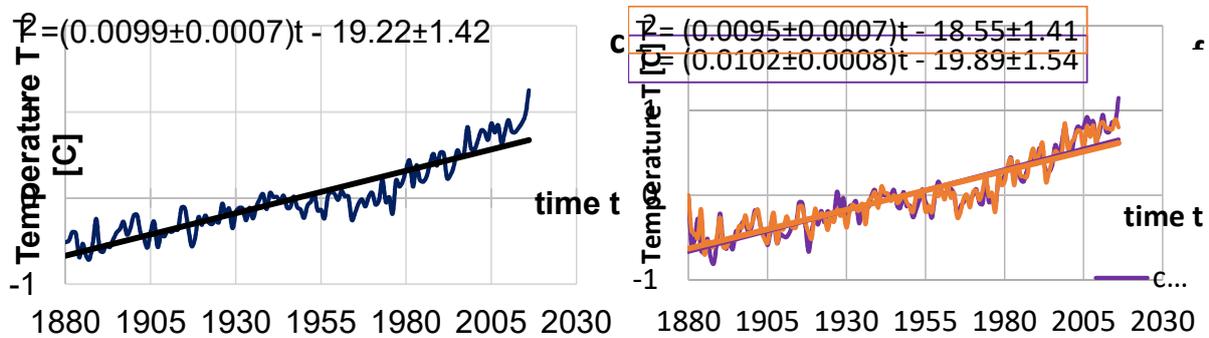

Figure 2a-f. (left panel) Annual changes of the average near-surface air temperature for Poland (a), Tbilisi (b) and the average global near-surface air temperature (c) and (right panel) for cold months (J-F-M-O-N-D) and hot (A-M-J-J-A-S) for Poland (d), Tbilisi (e) and average global temperature (f) in the period of 1881-2016 with linear regression lines.

It is worth mentioning that considered data series of T in Poland is available from 1781-2016. For Poland the centenary change in 1781-2016 is a bit less $\Delta T=\sim 0.7^0C$ (Figure 3a) than in 1881-2016 ($\Delta T=\sim 1.1^0C$). Figure 3b shows that the centenary change of the near-surface air temperature in Poland for cold season ($\Delta T=\sim 1^0C$) is ~twice larger than for summer season ($\Delta T=\sim 0.4^0C$). Figure 3c shows that data T in Poland has some peculiarity with almost stable average value of T for the period 1781-1880 ($\Delta T=\sim 0.1^0C$) and with increasing trend for the period 1881-2016 ($\Delta T=\sim 1.1^0C$), being comparable to Tbilisi and global changes of T ($\Delta T=\sim 1^0C$) in this period.

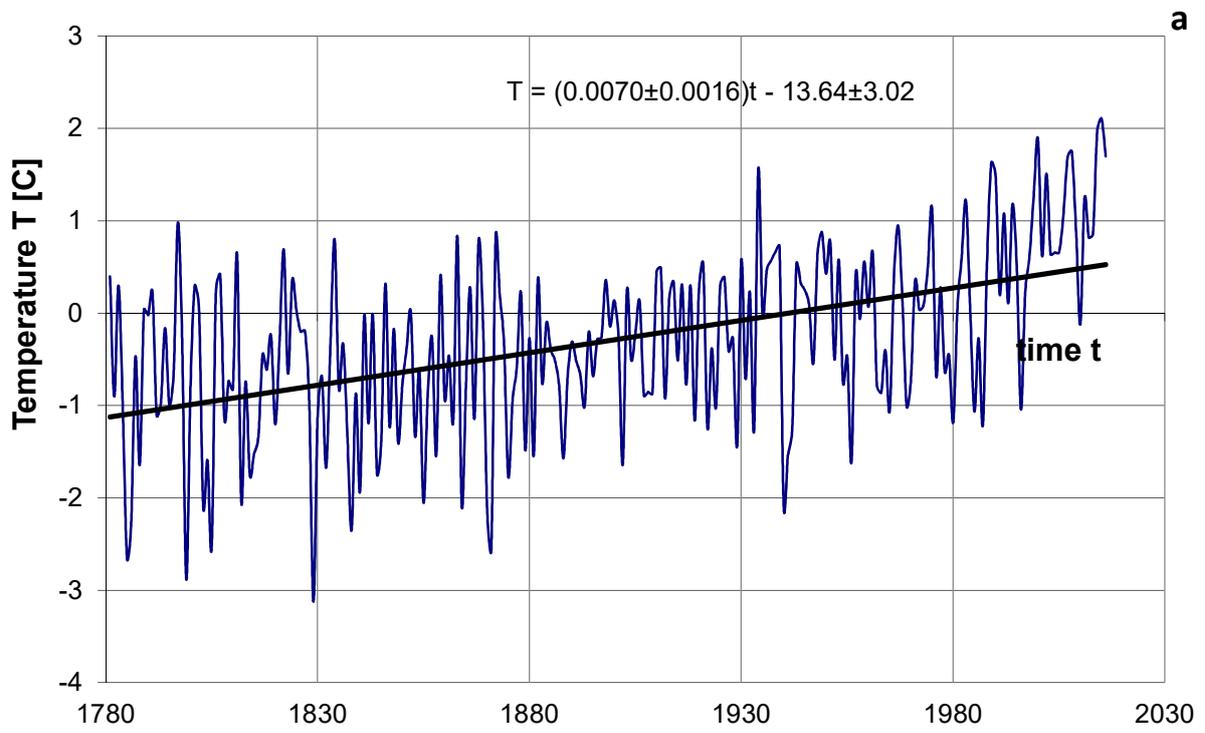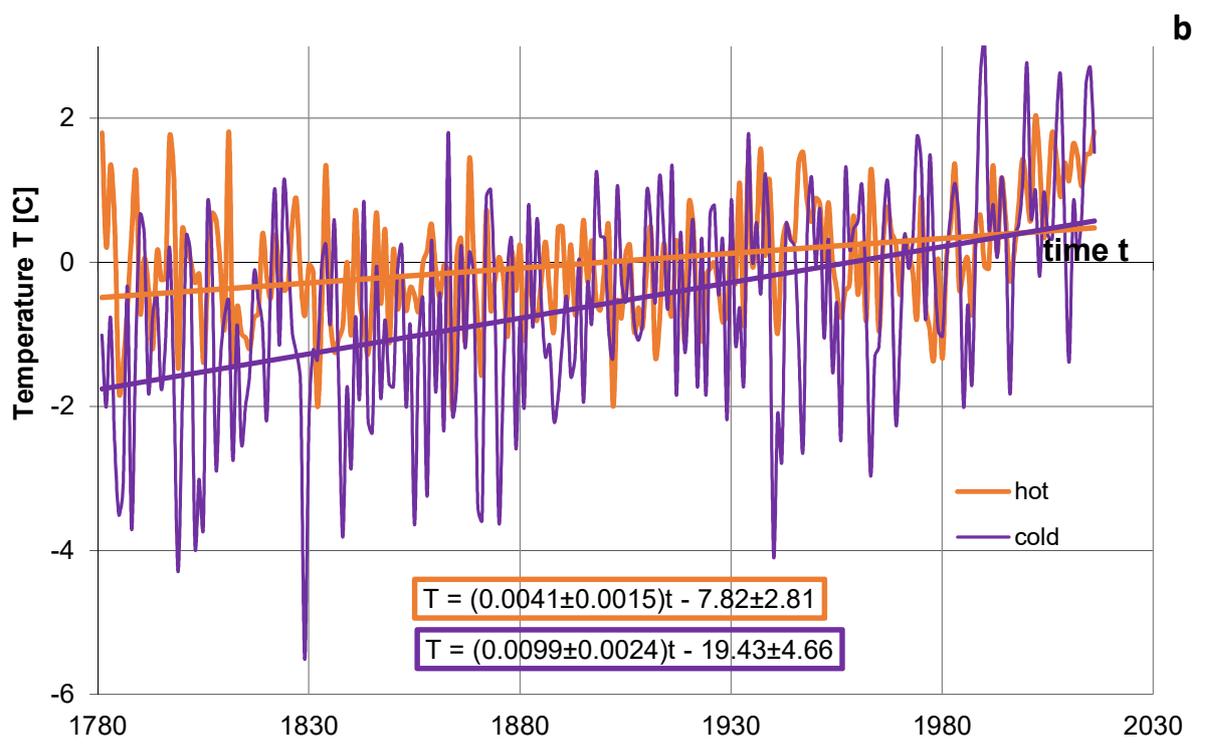

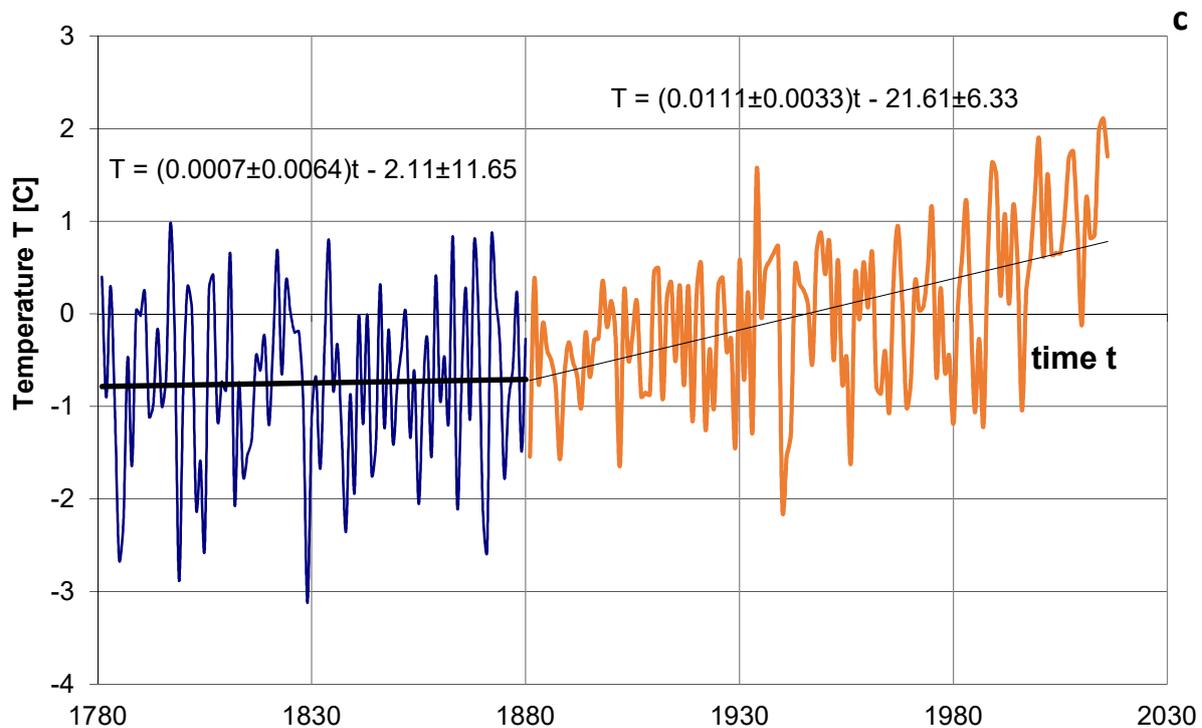

Figure 3abc. (a) Annual changes of the near-surface air temperature T in Poland for 1781-2016; (b) for cold months (J-F-M-O-N-D) and hot (A-M-J-J-A-S); (c) for 1781-1880 and 1881-2016 with corresponding regression lines.

3. On relationship of the near-surface air temperature and solar activity

It is of great interest to search what phenomenon or a group of phenomena single-valued are responsible for the centenary warming effect. There are few wide-ranging alternatives, among them – changes of solar activity and/or human activity. Nowadays, when reliable climate data is available for more than a century in many parts of the world, it is possible to seek a relationship between the solar activity and the climate. At present solar variability is accepted as a major contributor to climate change on a global scale in the pre-industrial period, as well as being probably an important factor even today [for a review see e.g. *Reid*, 2000; *Haigh*, 2007]. The effects of solar variability on terrestrial climate are discussed in several works [e.g., *Kirkby*, 2007; *Solanki et al.*, 2013; *Kudela*, 2013; *Gervais*, 2016; *Mouël et al.*, 2019; *Veretenenko and Ogurtsov*, 2020].
Unfortunately, it is difficult to call a clear quantitative global proxy to estimate a role of human activity in centenary warming, but to describe changes of solar activity one can employ a few parameters [e.g. *Lean et al*, 2002; *Friis-Chritensen and Larsen*, 1991], among them, e.g. very broadly used data of sunspot number SSN and total solar irradiance TSI.
Although SSN data is commonly used as an indicator of solar activity with more than 250 years history of observations, it is well known that SSN probably is not good signature of long term changes of real solar radiation [*Eddy*, 1976], therefore we decided to use for correlation analysis of the near-surface air temperature T in Poland and in Georgia with both SSN and TSI data.

As far in annual changes of SSN and TSI dominates the 11-year waviness, while in annual changes of T there is observed nothing like that. It is of interest, whether the centenary warming effect is related with the long period changes of solar activity. To reveal long period relationship between series of T&SSN and T& TSI data, the 11 year cycling changes were removed from all data using method of running smoothing. The 11-year periodicity is hardly recognizable in changes of T (Figure 1a), but to make data of SSN, TSI and T to be similarly homogeneous (for the sake of comparison) a running smoothing operation has carried out for all data.

Data of SSN and TSI show a clear quasi cycling changes [*Balogh et al.*, 2014; *Coddington et al.*, 2016] with dominated ~ 11 year periodicities, while T displays some fluctuations with a little non significant gradually increasing (positive drift) versus time t. At the first glance there is hardly recognizable any explicit relation between T&SSN confirmed by very small correlation coefficients (for global T r =0.09±0.01, for Poland r = 0.12±0.01 and for Tbilisi r = 0.14±0.01 in whole period of 1881-2016). For T&TSI correlation is relatively higher but still less than 0.5 (for global T r = 0.47±0.01, for Poland r = 0.33±0.01 and for Tbilisi r = 0.39±0.01 in whole period of 1881-2016). To show whether there exist relations between T&SSN and T&TSI on short time scales we calculate a correlation coefficient for short periods, as well.

In Figures 4abc are presented temporal changes of correlation coefficients r between annual T&SSN (blue), T&TSI (violet), 11-years smoothed data of T&SSN (red) and 11-years smoothed data of T&TSI (green) calculated for running 50 years span for Poland (Figure 4a), Tbilisi (Figure 4b) and global temperature T (Figure 4c) for 1911-1986. Unfortunately, to use data less that 50 years duration to calculate correlation coefficient is not statistically reliable. Figures 4(a)-(c) show that a correlation coefficient for original annual data is small (less than 0.5) for any 50 years time interval. So, one can state that between annual changes of T&SSN and T&TSI one could not recognize neither short nor long period noticeable relation. This results were expected, as far in annual changes of SSN and TSI dominates the 11-year waviness, while in annual changes of T this periodicity is not observed. A correlation coefficients r between changes of 11 years smoothed data of T&SSN (red lines in Figure 4) are valuable and have maximal values equal r= 0.66±0.07 for Poland, r= 0.82±0.05 for Tbilisi and r= 0.91±0.05 for global temperature T data from ~1890 up to ~1960. The maximal correlation coefficients r are even higher for 11 years smoothed data of T&TSI (green lines in Figure 4): r= 0.73±0.07 for Poland, r= 0.90±0.05 for Tbilisi and r= 0.95±0.05 for global temperature.

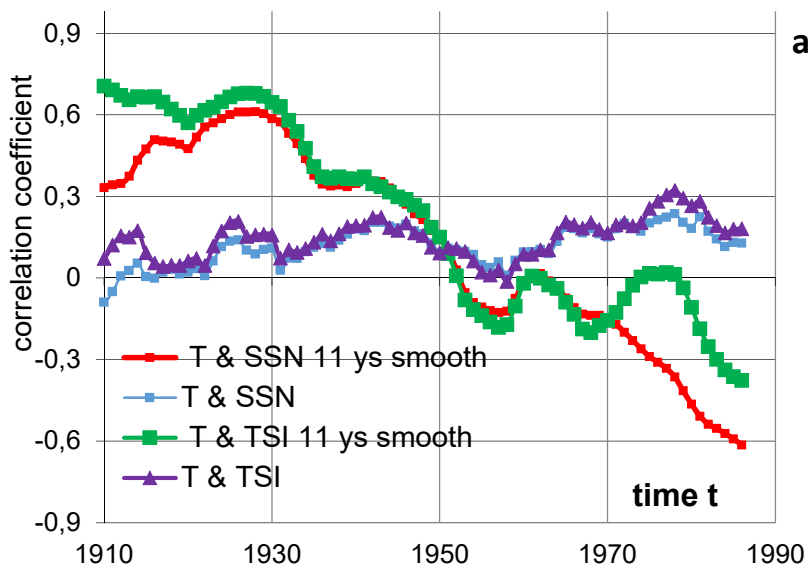

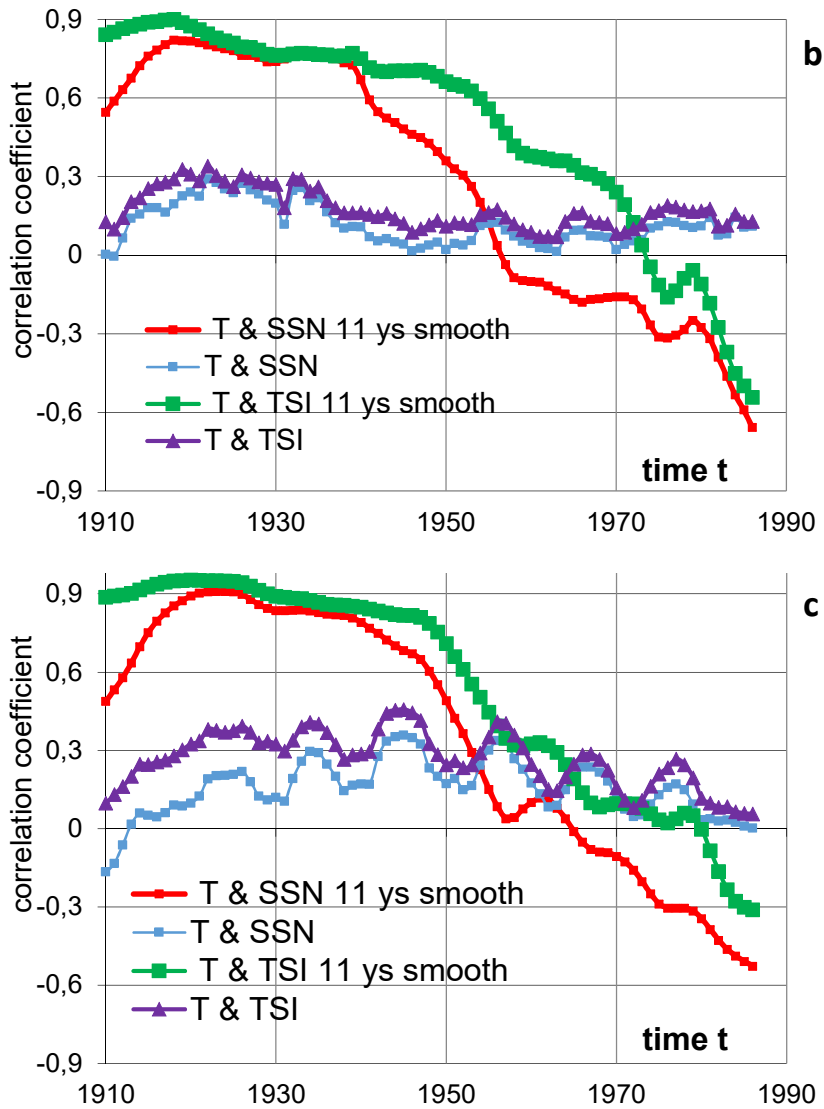

Figure 4abc. Temporal changes of the correlation coefficients between: annual near-surface air tempeature T and SSN (T&SSN, blue), T&SSN for 11-years smoothed data (red), annual near-surface air tempeature T and TSI (T&TSI, violet), T&TSI for 11-years smoothed data (green) for Poland (a), Tbilisi (b) and global (c) for 1911-1986.

However, in both T&SSN and T&TSI data there are observed some peculiar changes in time intervals less than the whole period to be analyzed. Figures 4abc show that a correlation coefficient is very high (r≥~0.6) since 1890 up to 1960, next a correlation coefficient decreases and becomes negative. It is clear that the whole period of 1881-2016 can be divided in two sub time intervals - the first 1890-1960 with a high correlation between changes of the T&SSN and T&TSI, and the second interval 1960-2016 with a weaker negative correlation between them. As far the TSI data better visualize the long period trend in solar activity, it even strengthen the statement that terrestial temperatures are strongly related with solar activity especially in 1890-1960. Thus, based on the correlation between T&SSN and T&TSI, a centenary warming effect in the period of 1890-1960 (~ during solar cycles 13-20) is mainly related to the change of solar activity. Similar results are shown for global T in [*Reid*, 1991; *Friis-Chritensen and Larsen*, 1991; *Zherebtsov, et al.*, 2019; *Veretenenko and Ogurtsov*, 2019]. Of course one could not exclude roles of other phenomena, among them human activities, but data of T in Tbilisi and Poland single-valued show that a decisive role in centenary changes (global warming) belongs to the changes of solar activity before years of sixties of last century (1890-1960). Indeed, after sixties of 20th century, when human activities were intensified, one can see gradual increase of global T, for Poland it happened a bit later after eighties, but for Tbilisi a warming effect remains on the zero level up to end of 20th century. However [*Laptuchov and Laptuchov*, 2010; 2015] indicates that in the long time history the solar activity rather than human activity plays the determining role in the observed climate warming, especially an increase in the global near-surface air temperature is accompanied by increase in the carbon dioxide amount in the atmosphere due to its evaporation from the seawater, not by human beings. Recently, [*Zherebtsov, et al.,* 2019] pointed out that the global warming in the 21st century has dramatically slowed down, almost stopped based on the sea surface temperature, whereas the concentration of the carbon dioxide in the atmosphere has been increasing, although solar and geomagnetic activities have noticeably weakened over the last decade. Moreover, in the sequence of papers [*Veretenenko and Ogurtsov*, 2018, 2019, 2020] show that the long term variations of total solar irradiance may be a possible cause of ~60-years oscillations of global temperature anomalies and the corresponding changes of the circulation regime in the lower and middle atmosphere. *Gervais* [2016] shows that changes of the air temperatures in the beginning of the 21$^{st}$ century indicate for the onset of the declining phase of the 60-year cycle, being in phase with the Atlantic Multidecadal Oscillation.

4. Study of short period fluctuations of the near-surface air temperature using spectral and wavelet analysis

The near-surface air temperature at Earth measured since ~1880 displays various quasi-periodic oscillations with different time scales [see e.g. *Baliunas et al.,* 1997; *Palus and Novotna*, 2004; *Palus,* 2014; *Francia et al.*, 2015 and references therein]. *Baliunas et al.,* [1997] analyzing the central England temperature T from 1659-1990 found the peaks at $7.5 \pm 1.0$ years, $14.4 \pm 1.0$ years, $23.5 \pm 2.0$ years, as well as long period variation at $102 \pm 15$ years. The most prominent nonrandom periodic variations reported by *Palus and Novotna* [2004] is the period around 7–8 years observed in the near-surface air temperature and other meteorological data. Considering geographical locations of the stations, the effect of 7-8 years oscillations was found in the stations located slightly over 50 degrees of northern latitude [*Palus*, 2014]. This author suggests that the observed phenomenon might be connected with local effects of the North Atlantic Oscillation, one of the global modes of the atmospheric circulation variability.
In this paper we concentrate on the relationship of the near-surface air temperature data T in Poland and in Tbilisi with solar activity. We analyze the quasi-periodic character of the temperature T as well. Identification of the quasi-periodic variation in temperature data was done by the wavelet and cross-wavelet methods. The wavelet technique [*Torrence and Compo*, 1998; *Grinsted et al.*, 2004] allows to study the dynamic of the hidden periodicity in the time-frequency space. We used the wavelet software available at the website [http://paos.colorado.edu/research/wavelets/software.html].
The Matlab package (http://noc.ac.uk/usingscience/crosswaveletwaveletcoherence) of the National Oceanography Centre, Liverpool, UK was used in the calculation of the cross-wavelet spectrum.
In this paper the cross-wavelet transform is used in the calculation of the degree of cause and effect dependence of the observed near-surface air temperature T data and solar activity.
First we study the period of high correlation between solar activity and T. Figure 5 demonstrates the 11-years smoothed data of temperature T for Poland (a) and Tbilisi (b) for the period 1885-2010. Figure 5 indicates peculiar quasi-periodic changes of T data for 1885-1980. This period averagely corresponds to the period of the highest correlation between solar activity and T. After removing linear trend from changes of T (Figure 5a for Poland and Figure 5b for Tbilisi) we analyze the quasi-periodic character using wavelet method. Results of calculation are presented in Figures 6ab. Figures 6ab show that there exists a quasi-periodic wave of the T for both series, for Tbilisi the main pick is at ~23 years, for Poland data the main picks are at ~8 years, also confirmed earlier for north stations by (*Palus*, 2014) and 17 years. Figures 6ab indicate that the temperature T data has some quasi-periodic character with a period of $20\pm3$ years. Most likely this quasi-periodic fluctuations of T can be ascribed to the 22-year solar magnetic cycles, related with the reversal of the Sun's global magnetic field taking place in maximum epochs of 11-year cycles of solar activity.

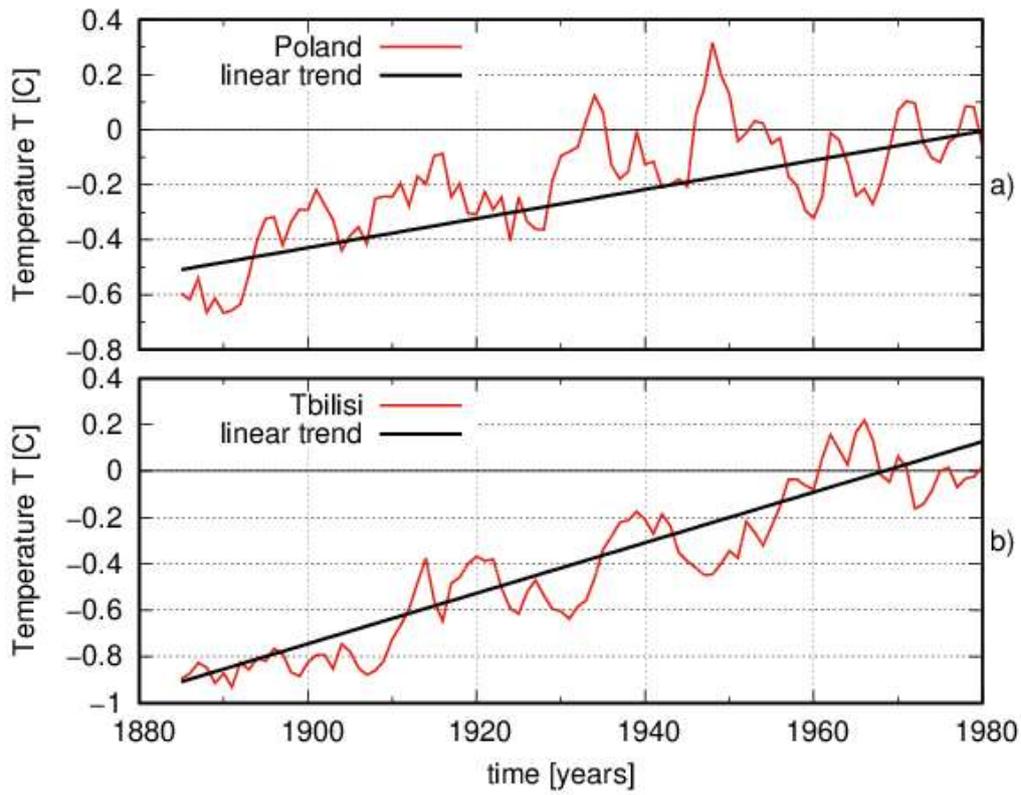

Figure 5ab. Smoothed over 11 years the near-surface air temperature T in (a) Poland and (b) Tbilisi for 1885-1980 with linear trend.

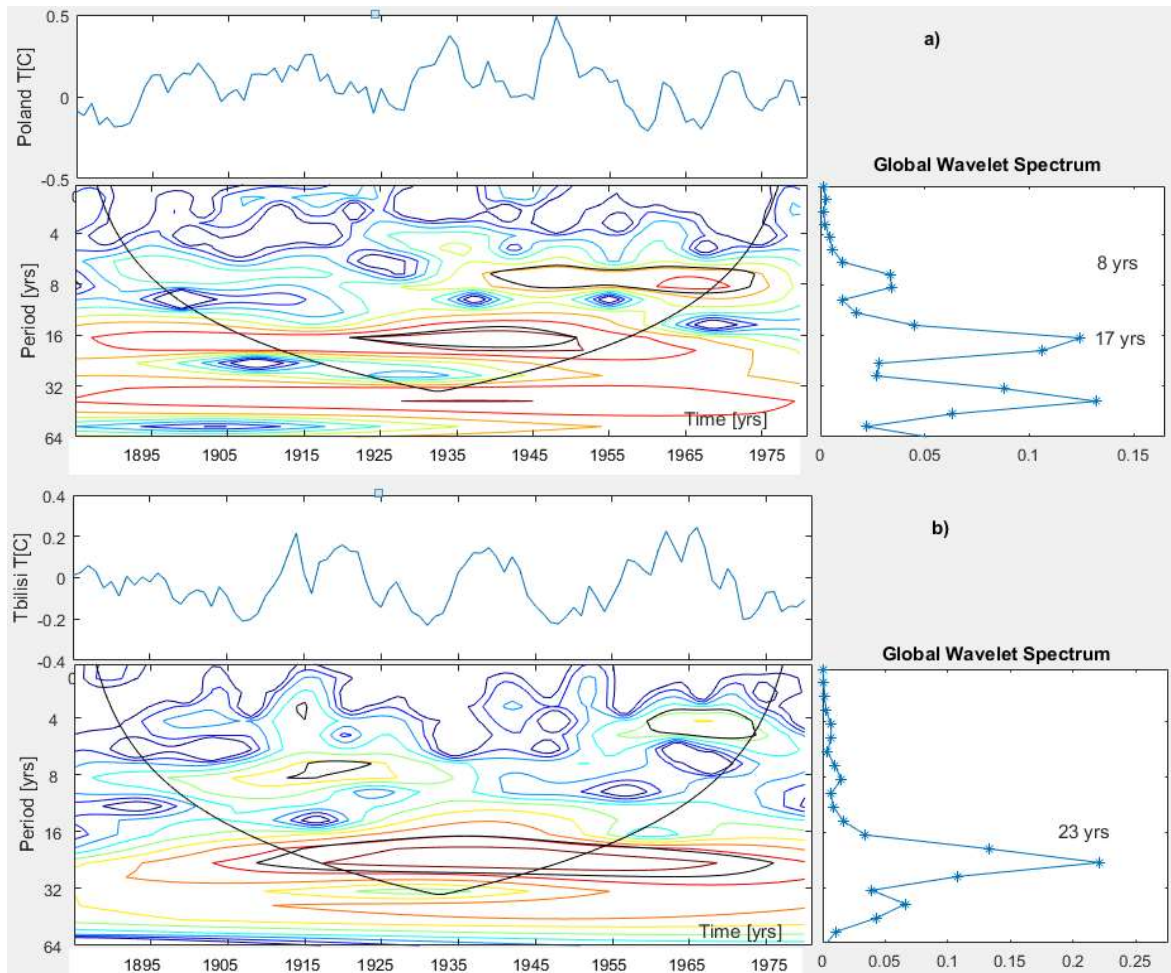

Figure 6ab. Wavelet spectrum of (a) Poland T, the main picks are ~8 and 17 years and (b) Tbilisi T, the main pick is ~23 years in 1885-1980.

## 5. Discussion and conclusions

We found an average centenary warming effect $\Delta T=(1.08 \pm 0.29)^0C$ based on the annual data of the near-surface air temperature T in Poland and in Tbilisi for period of 1881-2016. A centenary warming effect is larger in winter season ($\Delta T = \sim 1.15^0C$) than in other seasons ($\Delta T = \sim 0.95^0C$). We established that a centenary warming effect in Poland is almost insignificant ($\Delta T=\sim 0.1^0C$) in period 1781-1880, in comparison with 1881-2016. The time interval ~70 years (1890-1960) decisively contributes in creation of global warming effect, when a correlation coefficient between changes of solar activity and T is very high. The correlation coefficients between 11 years smoothed sunspot number SSN and T and between total solar irradiance TSI and T are: for Poland $r = 0.66 \pm 0.07$ and $r = 0.73 \pm 0.07$ and for Tbilisi $r = 0.82 \pm 0.05$ and $r = 0.90 \pm 0.05$, respectively. The centenary changes of temperature (global warming effect) generally should be related with the changes of solar activity. **Our aim is just to confirm that Poland and Tbilisi data show the same relation with solar activity as already existed in literature.** We demonstrated a few feeble $20 \pm 3$ year disturbances in changes of T by both - Tbilisi and Poland data for period 1885-1980, most likely related with the fluctuations of solar magnetic cycles. **Naturally for us this period might be connected with the polarity of the Sun, but we do not exclude other sources, e.g. being the higher harmonic of the multi-decadal and/or centennial changes of the long term variations of total solar irradiance being the possible cause of ~60-years oscillations of global temperature anomalies. We just confirm that Poland and Tbilisi data show the same pattern as observed e.g. in *Baliunas et al.,* papers.** Using the wavelet method we recognized some short period changes (7-8 years) in Poland near-surface air temperature T data, possibly potentially connected with local effects of the North Atlantic Oscillation, one of the global modes of the atmospheric circulation variability. **We have just observed this kind of periodicity. Additionally we confirm this period to be connected with the global modes of the atmospheric circulation following others works, e.g. papers by *Palus* and coauthors. This topic is still under discussion.**

Acknowledgments

For calculations we used the Matlab packages
[http://paos.colorado.edu/research/wavelets/software.html] and
[http://noc.ac.uk/usingscience/crosswaveletwaveletcoherence].
We use data of temperature in Poland https://meteomodel.pl/klimat/poltemp/ provided by The Institute of Meteorology and Water Management -National Research Institute (IMGW-PIB) (https://dane.imgw.pl), in Georgia data was provided by Institute of Hydrometeorology (Tbilisi).
We use data of solar activity: total solar irradiance http://lasp.colorado.edu/lisird/data/nrl2_tsi_P1Y/ and sunspot number http://www.sidc.be/silso/infossntotyearly.
**This paper is dedicated to the memory of Michael Alania who passed away on Monday, 18 May 2020 at age 85 years; he supervised RM's PhD degree work.**